\documentclass[amsmath,amssymb,aps,prl,amssymb,showpacs,nobibnotes
,twocolumn]{revtex4-2}
\usepackage{graphicx}
\usepackage{epstopdf}
\usepackage{hyperref} 
\usepackage{dsfont}    
\usepackage{color}

\newcounter{multieqs}

\newcommand{\be}{\begin{equation}}
\newcommand{\ee}{\end{equation}}
\newcommand{\eq}[1]{(\ref{#1})}
\newcommand{\bit}{\begin{itemize}}  \newcommand{\eit}{\end{itemize}}
\newcommand{\ben}{\begin{enumerate}}  \newcommand{\een}{\end{enumerate}}

\def\bea{\begin{eqnarray}}
\def\eea{\end{eqnarray}}
\let\bm=\bibitem

\def\la{\langle}
\def\ra{\rangle}
\def\div{\nabla \cdot}

\def\one{\mbox{1 \kern-.59em {\rm l}}}

\def\a{\alpha}        
\def\b{\beta}       
\def\g{\gamma}    
\def\d{\delta}  \def\D{\Delta}  
\def\e{\epsilon}
\def\ve{\varepsilon}

\def\l{\lambda} \def\L{\Lambda}
\def\m{\mu} \def\n{\nu}
\def\o{\omega}
 
\def\r{\rho}
\def\s{\sigma}  
\def\t{\tau}
\def\th{\theta}

\DeclareMathAlphabet\mathbfcal{OMS}{cmsy}{b}{n}
\def\bcJ{\mathbfcal{J}}
\def\cJ{\mathcal{J}}

\def\cS{{\cal S}}

\def\ps {{p \hspace{-6.4pt} \slash}\;}

\def\psib{\bar{\psi}}

\def\bp{{\bf p}}

\def\bv{{\bf v}}

\def\bE{{\bf E}}

\def\bS{{\bf \cS}}

\def\d{\delta}\def\D{\Delta}

 \def\del{\partial}

\def\uno{\mbox{1 \kern-.59em {\rm l}}}

\def\one{1\!\!1\,\,}

\newcommand{\tr}{\mbox{tr}}

\def\Box{\square}

\def\bcomment#1{}

\def \qs {q\hspace{-.53em}/\hspace{.15em}}

\def\IR{\relax{\rm I\kern-.18em R}}
\def \id {{\bf 1}}

\def\bJ{\pmb{J}}
\def\bB{\pmb{B}}
\def\bM{\pmb{M}}

\def\bn{\pmb{n}}

\newcommand{\rrr}[1]{\vskip 0.2cm \noindent{\it #1} ---}

\begin{document}
\title{ Induced Quantized Spin Current in Vacuum}
\author{Chong-Sun Chu}
\author{Chun-Hei Leung}
\affiliation{Department of Physics, National Tsing-Hua
  University, Hsinchu 30013, Taiwan
}
\affiliation{Center for Theory and Computation, National 
  Tsing-Hua University, Hsinchu 30013, Taiwan}

\begin{abstract}

  We uncover a fundamental effect of the QED vacuum in an 
  external electromagnetic (EM) field. We show that the 
  quantized vacuum of electrons is spin polarized by the 
  EM field and manifests as a vacuum spin current. An 
  experiment is proposed to measure the spin torque 
  exerted by the spin current by measuring the twisted 
  angle of the director axis of a nematic liquid crystal.
  

\end{abstract}

\maketitle


\rrr{Introduction.}
The quantum nature of vacuum is a fascinating place to look for novel
physical phenomena. For example, the large scale fluctuation of the
Universe is supposed to emerge from the quantum ripples of the near de
Sitter vacuum in the inflationary universe \cite{inf}. In high energy
physics, the metastability of the Higgs vacuum \cite{m1} has
nontrivial cosmological consequences and could provide a new
observational window to particle physics well beyond what collider
experiments can achieve.

Recently, it was found that the vacuum fluctuations of boundary QED
system in the presence of an external magnetic field result in a
magnetization current near its boundary
\cite{Chu:2018ksb,Chu:2018ntx}. This nontrivial electromagnetic
response of the vacuum is due to the electric charges carried by the
virtual electrons and positrons of the theory. As these quantum
fluctuations also carry spin, it is natural to ask if and how the
quantum fluctuation of the spin degrees of freedom of the vacuum would
manifest in observation.

In this Letter, we show that due to the renormalization effect of the
spin-orbit coupling of the electrons, the quantized vacuum can become
spin polarized in the presence of an applied EM field and results in a
spin current; see Eq. \eq{final}. The vector spin current is
orthogonal to the applied electric field and the observed spin
polarization. We also discuss briefly how it may be observed
experimentally using a setup of nematic liquid crystal.

\rrr{Definition of spin current.}
The transportation of the spin of electrons is described by a spin
current. Classically, the spin current is given by the 3-tensor
$J^{ij} = n v^i s^j$, where $v^i$ is the velocity of the electron,
$s^j$ is the spin, $n$ is the particle number density, and $i, j
=1,2,3$ denotes the spatial directions.  Quantum mechanically, the
spin polarization of electrons is represented by the Pauli matrices
$\s^i$ in the rest frame of the electron and a definition of the spin
current:
\be \label{rashba}
J^{\m i} = \frac{1}{2}(v^\m \s^i+ \s^i v^\m), \quad J^{0i} = \s^i
\ee
in terms of a symmetrization of the velocity and the spin has 
been proposed by Rashba \cite{rashba}. Although this works well 
for a nonrelativistic system, the generalization to the 
relativistic case requires the use of a relativistic spin 
operator. Such an operator was constructed long ago by Bargmann 
and Wigner \cite{wigner} for the free Dirac theory, and it reads
\be \label{BW}
T_{\rm BW}^i = \b \Sigma^i + \g^5  p^i /m, \quad T_{\rm BW}^0=   {\bf
  \Sigma}
\cdot {\bf p}/m,
\ee
where $\b = \g^0$, $\a^i = \b \g^i$, $\g^5 = i \g^0\g^1\g^2\g^3$ is
the chirality matrix and $\Sigma^i = (i/2) \ve^{ijk} \g^j \g^k$. We
will adopt the particle physics convention of metric signature
$(+,-,-,-)$ in this Letter. The Bergmann-Wigner spin operator $T^\mu$
is a Lorentz 4-vector \cite{good}, it acts as the generator of the
little group of the Poincare group. That this is so is because it is
indeed equivalent to the Pauli-Lubanski operator
\be \label{PS}
W^\m = -\frac{1}{2m }\ve^{\m\a\b\n} J_{\a\b} p_\n. 
\ee
This can be seen by noticing that the spin angular momentum generator
is given by $J_{\a\b} = (i/2) [\g_\a,\g_\b]$ in the spinor
representation. It is then easy to see that Eq. \eq{PS} is equal to
the spin operator Eq. \eq{BW} on shell.

In this Letter, we are interested in the renormalization phenomena of
the QED vacuum in the presence of an external electromagnetic field
$A^\m$. As the velocity is represented by $v^\m = \b \g^\m$ in the
Dirac theory, we propose the following definition of spin current in
quantum field theory:
    \be \label{JS}
        J_S^{\m \a}(x) := \bar{\psi}(x) S^{\m \a}
      \psi(x),
      \ee
      where
      \be \label{S-kernel}
S^{\m\a} :=\frac{1}{2} [\gamma^\m T^\a(x) + T^\a(x) \gamma^\m]
      \ee
      and  $T^\a$ is the covariant  Bergmann-Wigner
      spin operator in a background electromagnetic
      field: 
\be \label{BW-A}
T^i = \b \Sigma^i + \g^5 \frac{ \pi^i}{m}, \quad T^0=  \frac{1}{m}
{\bf \Sigma}
\cdot {\bf \pi}.
\ee
Here $\pi^i = p^i -e A^i$ is the covariant momentum and $e>0$ is the
magnitude of the electric charge. Similar definitions \cite{vernes,taya}
have been considered before. Our definition is better justified since 
we have adopted a symmetrization prescription such that Eq. \eq{JS}
reduces to the one
(Eq. \eq{rashba}) of Rashba in the nonrelativistic
limit. A further justification of our definition \eq{JS}
can be obtained by noticing that the current Eq. \eq{JS}
satisfies the conservation law
\be \label{con-eqn}
\del_\m J^{\m \a} = \frac{e}{m} \cS_\n F^{\n\a},
\ee
with a source term. Here  $\cS^\n$ is the density defined by
\be
\cS^i : = \psi^\dagger \Sigma^i \psi, \quad \cS^0 : = \psi^\dagger \g^5 \psi.
\ee
We note in passing that $J^{\m\a}$ is not a Noether current, but
Eq. \eq{con-eqn} is obtained from the fermion equation of motion. For
the spatial directions $\a =i$, Eq. \eq{con-eqn} gives explicitly
\be \label{con}
          \frac{\partial \r^i }{\partial t}+ \div{\bJ}^i =
        \frac{e}{m}
        \Big( (\bS \times\bB)^i + \cS^0 E^i \Big).
\ee
Here $\r^i:=J^{0i}$ and $\bJ^i:={\bf e}_k J^{ki}$. It is clear that
$\r^i$ gives the spin polarization density. As a result, $\bJ^i$ does
admit the correct interpretation as a current density for spin
polarization in the $i$th direction. We also note that $-\div{\bJ}^i$
and the EM terms on the right-hand side of Eq. \eq{con} can be
interpreted as the spin torque from the matters and the external EM
fields. Classically in a vacuum, $\cS^\mu$ vanishes and the spin
current respects the continuity equation
    \begin{equation}
        \frac{\partial\r^i }{\partial t}  + \div \bJ^i = 0.
    \label{CONTINUITY}
    \end{equation}
We will show in this Letter that, in the presence of a background
electromagnetic field, the spin current $\bJ^i$ and the spin density
$\r^i$ becomes nonzero due to the polarization effect of the
electromagnetic field on the quantum fluctuations of the
vacuum. Nevertheless the conservation law \eq{CONTINUITY} is still
satisfied at the quantum mechanical level.

We end this section with a couple of remarks.  (i) Properly speaking,
the current Eq. \eq{JS} measures the flow of {\it spin polarization}
and should be called a spin polarization current. We will follow the
common practice of the literature, e.g., Ref. \cite{rashba}, and refer
to it as the spin current. To get the {\it spin momentum current}, one
needs to multiply Eq. \eq{JS} with $\hbar/2$ of the spin angular
momentum of each fermion.  (ii) We note that in the nonrelativistic
limit, the magnetic dipole
coupling and the spin-orbit coupling  term in the Dirac Hamiltonian
  \be
  H_{E,B} = -\frac{e }{2m} {\bf \Sigma} \cdot \bB
  -\frac{e}{4m^2} {\bf \Sigma}\cdot \bE \times \bp
  \ee
can be written in the form
  \be \label{HJF}
H_{E,B} = -\frac{e}{8m} \ve_{\m\n\l\s}J^{\m\n} F^{\l\s}, 
\ee
where $J^{\m \n}$ here is the spin current
\eq{rashba} in the
nonrelativistic limit. The form \eq{HJF} shows clearly that the spin
current, at least in the nonrelativistic limit, couples to the EM
field strengths instead of the EM potentials like the electric current
$J^\m =-e \psib \g^\m \psi$.
(iii) Equation \eq{HJF} also shows that a certain amount of energy is
needed to generate the stated vacuum spin current and this is supplied
by the external power source that maintains the background EM field
configuration.
(iv) Finally, we remark that for a quantum field theory in curved
space, there is a current $J_\m^{ab}$ that couples to the spin
connection $\o_\m^{ab}$. Here $a,b = 0, 1, 2, 3$ refer to the frame
indices. This current couples to gravity and is also sometimes
referred to as a spin current in the respective community. However we
emphasize that this is different from the spin current we introduced
in this Letter.

    \rrr{Vacuum expectation of spin current.}
In quantum field theory, the spin current \eq{JS} is a composite
operator which needs to be renormalized. We are interested in the
vacuum expectation value (vev) of the spin current in a background
electromagnetic field $A_\m$.  This can be computed in perturbation
theory as
\be \label{J-vev-x}
\la J^{\m\a}(x) \ra_A = -i e \int d^4y \la   J^{\m\a}(x) J^\b(y)
\ra A_\b (y) + O(A^2),
\ee
or, it can be written in the momentum space as
\be \label{JA-q}
\la J^{\m\a} (q) \ra_A = e T^{\m\a\b}(q) A_\b (q) + O(A^2), 
\ee
where $T^{\m\a\b}(q)$ is the Green's function
\be
 T^{\m\a\b}(q) := -i \int d^4x  e^{i qx}\, \la  J^{\m\a}(x) J^\b(0) \ra.
 \ee
 At 1-loop, $T^{\m\a\b}(q)$ is given by
 \be
 T^{\m\a\b}(q) = -i \int \frac{d^4 p}{(2\pi)^4} (-1) \tr \left(
 S^{\m\a} \frac{i}{\ps -m} \g^\b \frac{i}{\ps+\qs-m} \right)
 \ee
 where $S^{\m\a}$ is given by Eq. \eq{S-kernel}. The trace of the
 gamma matrices can be simplified and we obtain
 \be
T^{\m 0 \b} = 0, \qquad T^{\m i \b} = -4 m \ve^{i\m \d\b}  q_\d \, I(q), 
 \quad i =1,2,3,
\ee
where $I(q)$ is the momentum function defined by
\be \label{I}
I(q) : = \int  \frac{d^4 p}{(2\pi)^4} \frac{1}{p^2-m^2} \frac{1}{(p+q)^2 -m^2}.
\ee
Note that $I(q)$ and hence the Green's function is logarithmic
divergent. This is due to the singular product of local quantum fields
in the expression \eq{JS} for the spin current operator. In order to
give a proper definition of this composite operator, one need to
isolate the divergent terms with a regularization scheme and subtract
it away with counterterms in the Lagrangian
\cite{renorm}. Regularizing Eq. \eq{I}, we obtain
\be \label{I1}
I(q) = \frac{i}{16\pi^2} 
\Big( \log \frac{\L^2}{\m^2} -1
+ \log \frac{\mu^2}{m^2} - h\Big(\frac{q^2}{m^2}\Big)+\cdots
\Big)
\ee
for a momentum cutoff regularization, and
\be \label{I2}
I(q) = \frac{i}{16\pi^2} 
\Big(\frac{2}{\e} - \g_E +\log \frac{4\pi\mu^2}{m^2} -h\Big(\frac{q^2}{m^2}\Big) 
+\cdots\Big)
\ee
for dimensional regularization to $d=4-\e$ dimension. Here $\cdots$
denotes terms of order $O(1/\L^2)$ or $O(\e)$, and $h(x)$ is the
function
\be
h(x):= \int_0^1 d\xi \, \ln[1+\xi(1-\xi)x]
\ee
with $h(0) =0$. We can now subtract away the divergence with a counterterm
and obtain 
\be  \label{vev-renorm-J-k}
    \la J_R^{\m i}(x) \ra_A = -\frac{e m c}{8 \pi^2 \hbar}
   \ve^{i \m \d  \b} \Big( \log\frac{m^2}{\m^2} +a_J
   + h\Big(\frac{\Box}{m^2}\Big)\Big) F_{\d \b}(x)
   \ee
and $\la J_R^{\m 0}(x) \ra_A =0$. Here $\m$ is the standard RG scale of the
QFT and $a_J$ is an independent arbitrary constant that is due to the
arbitrariness in the choice of the finite part of the counterterm for
the composite operator $J^{\m i}$.

To uniquely fix the finite part (i.e., fixing $a_J$) and hence the
definition of the renormalized spin current, a normalization condition
is required. Physically, if the electron mass is sent to infinity,
then the quantum loop effects are completely suppressed and the
renormalized spin current mush vanish. In a theory with a cutoff $\L$,
the fermion loop effects are now suppressed as $m$ approaches this
scale. As a result, we have the decoupling condition
\be \label{phys-cond}
\lim_{m \to \L} \la J_R^{\mu i} \ra_A = O(\frac{1}{\L}).
\ee
This serves as a natural normalization condition for the renormalized
spin current. However extra care is needed for QED where there is a
Landau pole and the UV cutoff cannot be taken to exceed that. In fact,
the positiveness of the beta function $\b (\a) = 2\a^2/3\pi$ ($\a=
e^2/4\pi \hbar c$ is the fine structure constant) give rises to the RG
flow for the coupling
\be
\frac{1}{\a(\m)} - \frac{1}{\a(\m_0)} = -\frac{2}{3\pi} \log\frac{\m}{\m_0}.
\ee
This implies the presence of a cutoff scale $\L_L = m e^{3\pi/2\a}$,
the Landau scale, where the bare coupling $\a(\L_L)$ becomes
infinite. The Landau scale represents the highest possible UV cutoff
one may utilize in the renormalization program of QED. Taking
$\L=\L_L$ in Eq. \eq{phys-cond}, the renormalization constant $a_J = -
\log \L_L^2/\mu^2$ is fixed up to $O(1/\L_L)$ terms, which we will
ignore. Restoring the units of $c/\hbar^2$ \cite{dim}, we finally
obtain 
\be \label{final}
  \la J_R^{\m i}(x) \ra_A = \frac{e m c}{8 \pi^2 \hbar^2}
   \Big(\frac{3 \pi}{\a} -h\Big(\frac{\Box}{m^2}\Big)\Big) \ve^{i \m \d  \b} F_{\d\b}.
   \ee
Note that the current Eq. \eq{final} is conserved
due to the Bianchi identity of the EM field. For a classical electromagnetic
background in vacuum, it is $\Box F_{\a\b} =0$ and we have
\be \label{vev-classical}
\la J_R^{\m i}(x) \ra_A = \frac{3 emc}{8 \pi \hbar^2 \a}
\ve^{i \m \d  \b} F_{\d\b}
   \ee
or, in terms of components explicitly
\begin{subequations}
   \label{jejb}
\bea
\la J^{ji}_R \ra_A &=& -\frac{3emc}{4\pi \hbar^2 \a}\ve^{jik}E^k ,
\label{jejb1}\\
\la J^{0i}_R \ra_A &=&  \frac{3emc}{4 \pi \hbar^2 \a} B^i \label{jejb2}
\eea
\end{subequations}

The results \eq{final}, \eq{vev-classical}, \eq{jejb} are the main
results of this Letter. For the rest of the Letter, we will be
focusing on the case \eq{vev-classical} of a classical
background. However for generality we present the result \eq{final} to
cover the situation where the background EM fields go beyond the
Maxwell description; for example, if quantum nonlinear corrections of
QED is included, or if the classical EM fields are coupled to other
background fields (e.g., an axion background).

A couple of remarks are in order. 
(i) Note that our result is independent of the adopted regularization
scheme. Apart from the cutoff regularization and the dimensional
regularization, one can also do the heat-kernel regularization and
obtain the same intermediate result \eq{vev-renorm-J-k}, and the final
result \eq{final} after imposing the normalization condition.
(ii) It is interesting that the normalization condition \eq{phys-cond}
fixes $a_J$ in terms of the renormalization scale $\mu$. In the end,
$\mu$ remains free in the QFT and it is remarkable that the
renormalized spin current is independent of it.
(iii) We remark that the renormalization condition \eq{phys-cond} is
essentially nonperturbative in nature as the resulting prediction
\eq{final} scales inversely with $\a$ and is not smoothly connected
with the free theory result.

It is instructive to compare our effect with the famous Schwinger
effect \cite{sch1,sch2}, which refers to the production of
electron-positron pairs under the influence of an applied electric
field. The Schwinger effect is nonperturbative and requires a strong
electric field stronger than the critical field strength $E \gtrsim
E_{\rm critical} = m^2 c^3/(e \hbar)$ in order to produce an
observable amount of particle pairs. Unlike the Schwinger effect,
the spin current predicted in this Letter is a
nonperturbative consequence of
the nontrivial spin polarization of the vacuum. There is no real
production of particles involved. Our result is also different from
the analysis of Ref. \cite{taya} where the Schwinger-effect-produced
electric current is acted on by a second electric field to produce a
spin current. In this case, the spin current is generated in a similar
manner as in an ordinary material sample except that the source
electric current has a nonperturbative origin and so very small in
magnitude.

In the above we have considered pure QED. In a more realistic setting
where QED is embedded as the low energy part of a consistent high
energy theory, e.g., a grand unified theory or string theory when
gravity is included, the Landau pole would be replaced by the
corresponding GUT or Planck scale. In this case, the 1-loop results
\eq{final} will have the factor $3\pi/\a$ replaced by $3\pi(1/\a -
1/\a_G)$ where $\a_G$ is the fine structure coupling at the
unification scale. It is interesting that the vacuum expectation value
of the spin current actually provides a probe to the UV physics. The
value of $1/\a_G$ is model dependent. For example, for the MSSM GUT,
one has \cite{pdg2020} $1/\a_G \simeq 24.3$ and $M_G\simeq 2 \times
10^{16}$ GeV. In any case, $1/\a_G$ is expected to be small compared
to the observed $1/\a$ at the electron mass scale. In the following,
we will continue to analyze the result \eq{final} for pure QED, but
keeping in mind the overall magnitude of the spin current may be
different from Eq. \eq{final} by a small fraction.

\rrr{Physical picture.}
It may appear strange that the switching on of an electric field in
vacuum can produce an observable spin current \eq{jejb}. However the
physical origin of the spin current can be easily understood in turns
of the spin-orbit coupling of the quantum fluctuation of the
vacuum. To see this, let us consider the nonrelativistic expansion of
Dirac’s Hamiltonian up to $O(p^4)$ where a spin-orbit coupling term
arises,
$H_\text{SO} = -[(e \hbar)/(4 m^2 c^2)] {\bf \Sigma}
     \cdot   \bE \times \bp$. 
It is well known that the spin-orbit coupling term allows impurity in
a material to scatter the electrons in a spin-dependent way (skew
scattering) and generates a spin current \cite{smit1,smit2}. The
spin-orbit coupling also give rises to a side jump
\cite{berger1,berger2} described by the additional velocity
    \be \label{vso}
    \bv_\text{SO} = \frac{\partial
      H_\text{SO}}{\partial \bp} = -\l\,
       {\bf \Sigma} \times \bE, \quad \l := \frac{e \hbar}{4 m^2 c^2},
       \ee
which produces a spin-dependent shift to the trajectory of the
electrons and contributes to the spin current.

In vacuum, there are no charge carriers to start with, but virtual
pairs of electrons and positrons can be created and live for a short
time before they annihilate back. In standard QFT without background
field, such vacuum polarization processes give rise to the running of
couplings and scaling of dimensions in the theory. However, new
observable effects may result in the presence of a background
field. Without an electric field, the created electron-positron pair
will move away from each other with opposite velocities $\pm u_{0}^j$
due to momentum conservation. When an $E$ field is turned on, the
particles acquire an additional velocity \eq{vso} in accordance with
their spin states. The velocities of the virtual $e^\pm$ particles are
given by $ u^j = \pm u_{0}^j + v_\pm^j$, where $v_\pm^j = -\l
\ve^{jkl} \s_\pm^k E^l$ and $\s^i_\pm$ are the spin operators for the
$e^+$ and $e^-$ respectively. Take a measurement of the spin in the
$i$th direction, the expectation value of the $j$th component of the
spin current
    $\la J^{ji} \ra =
    \la n \s_+^i (u_0^j+ v_+^j)\ra +
    \la n \s_-^i (- u_0^j + v_-^j)\ra$
    gives
    \be \label{jj}
    \la J^{ji} \ra = -  2 n \l \ve^{jik} E_k,
    \ee
where $n $ is the number density of the virtual pairs. In deriving
Eq. \eq{jj}, we have used $\la \s^i_\pm\ra =0$ and the correlations
$\la \s_\pm^i \s_\pm^j \ra = \d^{ij} \id$ for the vacuum. As a result,
the free field velocity parts in Eq. \eq{jj} make no contribution,
while the anomalous velocity part of the virtual $e^\pm$ contribute
equally to the spin current. With the estimate that there is one
virtual pair within each volume of Compton wavelength, $n \sim
(mc/\hbar)^3$, we recover the result \eq{jejb1} up to a numerical
coefficient (including the sign) of order 10. It is clear from this
quantum mechanical argument that the induced spin current is a result
of the nontrivial spin-spin correlation imprinted on the vacuum due to
the applied electric field. This is similar to the induced current
\cite{Chu:2018ksb,Chu:2018ntx} and the induced Fermi condensate
\cite{Chu:2020mwx,Chu:2020gwq} which are due to nontrivial
magnetization and condensation of the quantized vacuum as a result of
external fields.

\begin{figure}[t]
 \includegraphics[width=0.4\textwidth]{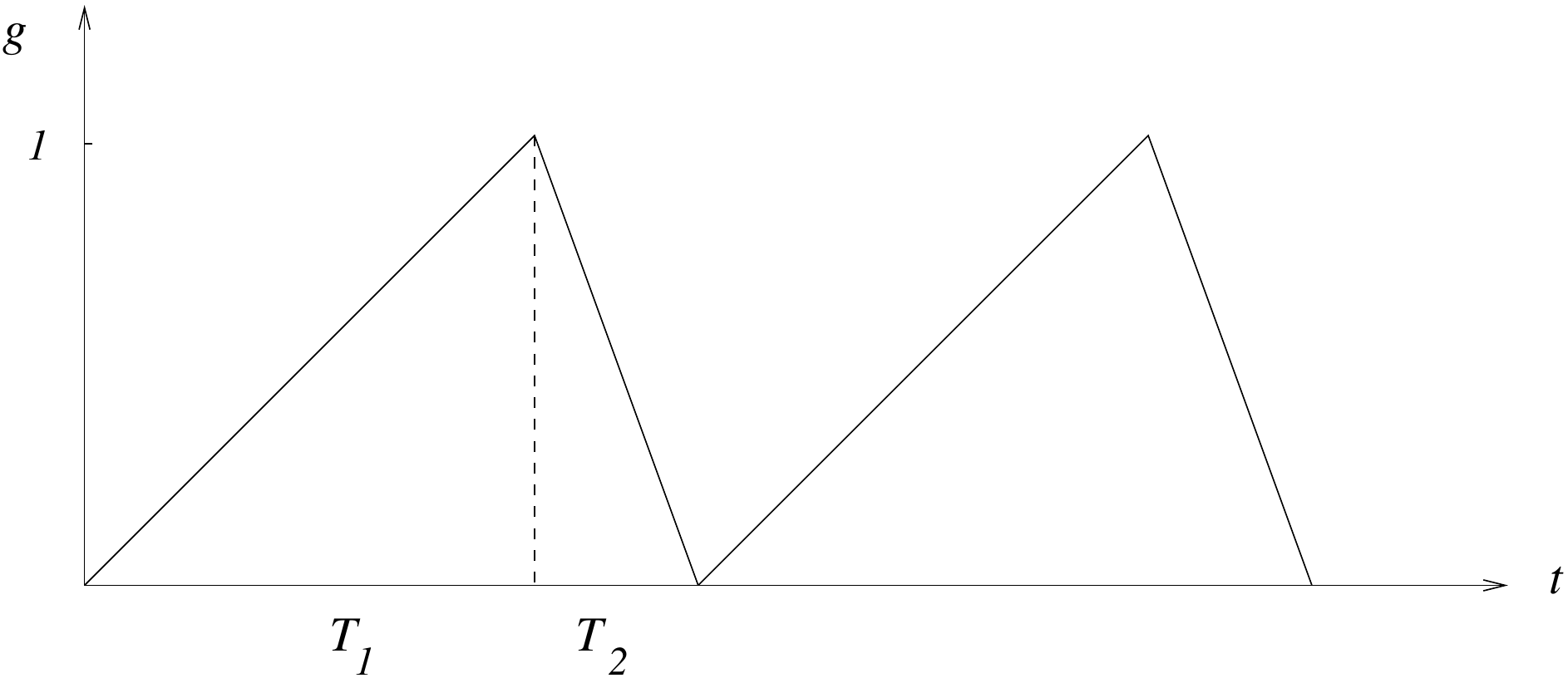}
\caption{Waveform of the $B$ field.}
\label{waveform}
\end{figure}

\rrr{Proposed experiment.}
The induced spin current may be observed by measuring the
torque exerted by the spin current on a probe placed in the vacuum.
Consider an infinitesimal volume element $\d V = \d x \d y \d z$ in the
interior of a probe placed under the influence of an external EM field. 
The spin momentum torque acting on $\d V$ is $\tau^i = - \hbar/2
\int_{\d V} d^3 x \, \div \la {\bJ}^i \ra $. Using the result
\eq{jejb} for a classical EM field in vacuum, we have
\be \label{SI}
\tau^i 
=  \frac{3emc}{8 \pi \hbar \a} \overline{\dot B^i} \d V,
\ee
where $\overline{\dot B^i} := \d V^{-1} \int _{\d V} \del B^i/\del t$
is the average rate of change of the magnetic field over the volume
$\d V$ We will be using the SI units from now on and hence a factor of
$c$ appears in Eq. \eq{SI}. For a $B$ field pointing in the $z$
direction described by a wave of the form
\be
B_z = B_0 g\Big(t-\frac{x}{c}\Big),
\ee
where $g$ is as in Fig. \ref{waveform},
we obtain
\be \label{spin-torque}
\tau^z=\frac{3emc}{8 \pi \hbar \a} B_0 f  \d V,
\ee
where $f:=1/T_1 -1/T_2$. For an external field with
$B_0 =100$ G, 
$f=1000$ Hz, the quantum spin torque per unit volume is
\be \label{tiny}
\tau^z/\d V = 6.5 \times 10^{-5} \, {\rm N m}{}^{-2}.
\ee
Not all of the vacuum torque is transferred to the probe. Physically
the angular momentum density $\cJ^i_0 = (\hbar/2) \r^i$ acquired by
the vacuum generates a vacuum magnetization $\bM_0 = (e/m) \bcJ_0$ and
this corresponds to quantum addition to the $B$ field, $\Delta\bB =
\mu_0 \bM_0$. For a material probe with magnetic susceptibility
$\chi$, the $B$ field generates a probe magnetization $\bM = \chi
\bM_0/(1+\chi)$. These are just the Barnett effect and the Einstein-de
Haas effect for the interplay between angular momentum and
magnetization \cite{landau}. This implies the probe receives a torque
$\tau^i_p$:
\be \label{tau-p}
\t_p^i = \eta \t^i,
\ee
where $\eta =4\pi \chi/(1+\chi) \approx \chi$ as $|\chi| \ll 1$. The
spin torque on the probe is thus tiny. Nevertheless, the twisting
effect may be observable by using a liquid crystal which is known to
be exceptional in sensitivity for torque measurement. In fact,
recently the Casimir torque exerted on the surface of a liquid crystal
\cite{somers1} has been observed successfully \cite{somers2}. We
propose here a similar setup to observe the spin torque arising from
the spin current.

\begin{figure}[t]
\includegraphics[width=0.45\textwidth]{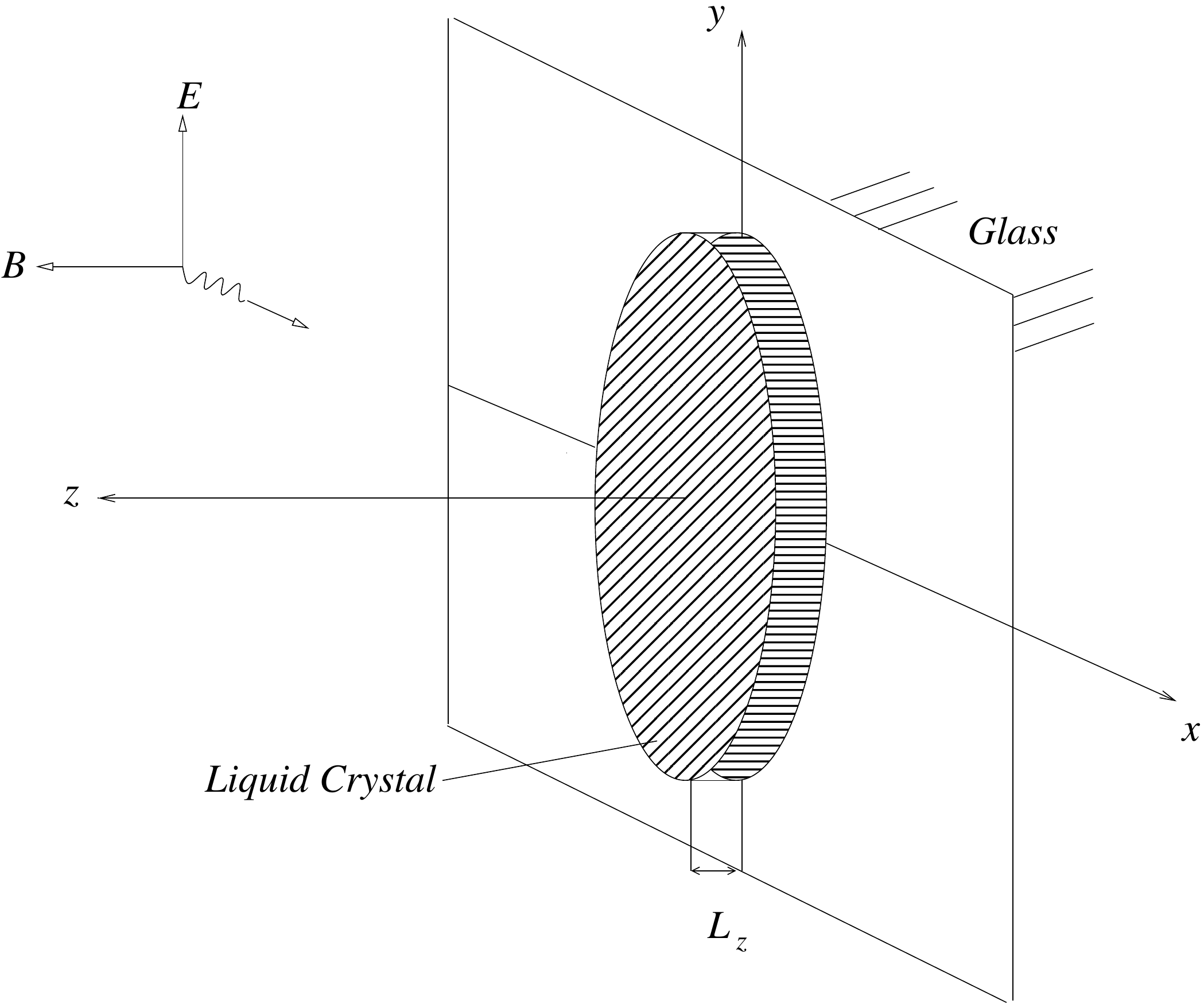}
\caption{ The director axis of a liquid crystal probe
  is twisted by the induced spin torque exerted by the vacuum.}
\label{LC}
\end{figure}
Consider a layer of nematic liquid crystal, e.g.,
4-cyano-40-pentylbiphenyl (5CB), placed in a vacuum with one side
($z=0$) anchored at a glass interface and the rest of the liquid
crystal ($0< z \leq L_z$) sits in the vacuum cavity, influenced by the
external EM field; See Fig. \ref{LC}. We place the liquid crystal such
that its director axis is always in the $xy$ plane, i.e., $\bn (z) =
[\cos \th(z), \sin \th(z), 0]$ where $\theta(z)$ describes the
orientation of the liquid crystal molecules with respect to the $x$
axis. Because of the action of the spin torque \eq{spin-torque}, the
molecules try to orient themselves correspondingly. An equilibrium
configuration is attained when this torque is balanced by the
restoring elastic torque of the liquid crystal.

Suppose there is no bend or splay of the liquid crystal, then only the
twist contribute to distorting energy density $u_d= (k/2) [(\del
  \th/\del z)]^2$, where $k$ is the twist elastic constant. For
example, $k= 3.6$ pN for the crystal 5CB.  The elastic energy stored
in $\d V$ is thus
\be
E_d [\th(z)] = A \int_z^{z+\d z}
dz
\frac{k}{2} \Big(\frac{\del \th}{\del z}\Big)^2,
\ee
where $A=\d x \d y$ is the area of the element.
This give rises to a restoring torque 
\be\label{t-bulk}
\tau_d(z) = \frac{\d E_d}{\d \th(z)} = -k \frac{\del^2 \th}{\del z^2} \d V.
\ee
This is in analogy with the Newton's second law $F = m \ddot{x}$ for
the inertia.  The equilibrium is attained when Eq. \eq{tau-p} is
balanced out by the restoring torque \cite{diff}. Note that the volume
factor cancels out. This gives
\be \label{thz}
\th(z) = \frac{\beta}{2} z^2, \quad \beta := \frac{3\pi\chi}{2\a}
\frac{m f_B f}{k \l_C},
\ee
where $f_B:= e B_0/(2\pi m)$ is the cyclotron frequency and $\l_C$ is
the Compton wavelength. In deriving Eq. \eq{thz}, we have taken the
boundary conditions $\th = \del_z \th =0$ at the glass contact $z=0$.

To enhance the detection of the twisted angle, a thicker liquid
crystal layer and a stronger magnetic field is preferred. However,
both of these are limited by the properties of the liquid crystal. In
order for the liquid crystal to be able to register the torque, the
response time of the liquid crystal should be smaller than $T_1$ and
$T_2$. This means $f$ is upper bounded by the response frequency
$f_r$: $f < f_r$. However, the response frequency is limited by the
thickness as generally the response time of the liquid crystal
increases with the thickness, linearly or quadratically depending on
the voltage \cite{lc}. As for the EM field, the field strengths cannot
be too strong as otherwise the crystal may be driven into a Freederick
transition \cite{lc}. Nematic liquid crystals are typically
diamagnetic with $\chi$ of the order of $10^{-5}$
\cite{lc}. Commercial nematic liquid crystal has a $f_r$ in the range
of 100 Hz and thickness of the order of 10 $\m$m. As an estimate,
consider a setup with a 100 G $B$ field and a nematic liquid crystal
with an elastic constant $k$ = 3.6 pN, thickness $L_z$ = 1 mm, a
response frequency of 1 kHz. The total twist angle accumulated over
the thickness of the probe is
\be
\D \th = 0.005^{\circ} \cdot
\frac{\chi}{10^{-5}}  \cdot
 \frac{3.6 \text{ pN}}{k} \cdot
 \frac{B_0}{100 \text{ G}}  \cdot
 \frac{f}{1 \text{ kHz}} \cdot
 \left(\frac{L_z}{1 \text{ mm}} \right)^2. 
\ee
This looks feasible. It will be interesting to perform an experiment
to make observation of the
spin current predicted in this
Letter.
The quantum spin current could have a wide range of
applications, from novel theoretical properties
of physical system, e.g. dark energy, to
practical effects on the workings of
micro-machined device.

\vskip 0.2cm
\acknowledgments
We acknowledge support of this work by the Grant
No. 107-2119-M-007-014-MY3 of the Ministry of Science and Technology
of Taiwan.

%


\end{document}